\documentclass[twocolumn]{aastex631}
\usepackage{amsmath}
\usepackage{booktabs}
\usepackage[utf8]{inputenc}
\usepackage{xcolor}
\usepackage{graphicx}
\usepackage{hyperref}
\usepackage{xspace}
\usepackage{orcidlink}
\usepackage{tabularx}
\usepackage{multirow}
\usepackage{array}
\usepackage{acronym}
\usepackage{capt-of}
\usepackage{etoolbox} 
\newcommand{\AckText}{}
\newcommand{\AddAck}[1]{\appto{\AckText}{#1}}

\newcommand{\tabref}[1]{Table~\ref{#1}}
\renewcommand{\eqref}[1]{Eq.~(\ref{#1})}

\newcommand{\figref}[1]{Figure~\ref{#1}}
\newcommand{\figsref}[2]{Figures~\ref{#1} and \ref{#2}}
\newcommand{\secref}[1]{Section~\ref{#1}}
\newcommand{\refref}[1]{\citet{#1}}
\newcommand{\appref}[1]{Appendix~\ref{#1}}

\begin{document}

\title{Fast targeted gravitational-wave followup search for compact binary mergers using \GSTLAL pipeline}

\author[0000-0003-0596-5648]{Leo Tsukada}
\email{leo.tsukada@ligo.org}
\affiliation{Department of Physics and Astronomy, University of Nevada, Las Vegas, 4505 South Maryland Parkway, Las Vegas, NV 89154, USA}
\affiliation{Nevada Center for Astrophysics, University of Nevada, Las Vegas, NV 89154, USA}
\AddAck{
LT acknowledges NASA 80NSSC23M0104 and the Nevada Center for Astrophysics for support.}

\author[0009-0003-3361-5538]{Noah Zhang}
\affiliation{School of Physics, Georgia Institute of Technology, Atlanta, GA 30332, USA}

\AddAck{
NZ and SS acknowledge support from the School of Physics, Georgia Tech and NSF Grant PHY-2409758.}

\author[0000-0002-0525-2317]{Surabhi Sachdev}
\affiliation{School of Physics, Georgia Institute of Technology, Atlanta, GA 30332, USA}
\author[0009-0004-2101-5428]{Shomik Adhicary}
\affiliation{Department of Physics, The Pennsylvania State University, University Park, PA 16802, USA}
\affiliation{Institute for Gravitation and the Cosmos, The Pennsylvania State University, University Park, PA 16802, USA}

\author{Chad Hanna}
\affiliation{Department of Physics, The Pennsylvania State University, University Park, PA 16802, USA}
\affiliation{Institute for Gravitation and the Cosmos, The Pennsylvania State University, University Park, PA 16802, USA}
\affiliation{Department of Astronomy and Astrophysics, The Pennsylvania State University, University Park, PA 16802, USA}
\affiliation{Institute for Computational and Data Sciences, The Pennsylvania State University, University Park, PA 16802, USA}
\AddAck{
CH Acknowledges generous support from the Eberly College of Science, the
Department of Physics, the Institute for Gravitation and the Cosmos, the
Institute for Computational and Data Sciences, and the Freed Early Career Professorship, supported by National Science Foundation Grants PHY-2308881 and PHY-2103662. }

\author[0000-0002-4148-4932]{Prathamesh Joshi}
\affiliation{School of Physics, Georgia Institute of Technology, Atlanta, GA 30332, USA}

\author[0000-0001-9675-4584]{Divya Singh}
\affiliation{Department of Physics, University of California, Berkeley, CA 94720, USA}
\AddAck{DS acknowledges support from NSF Grant PHY-2020275(Network for Neutrinos, Nuclear Astrophysics, and Symmetries (N3AS)). }

\begin{abstract}

We present a novel method to conduct targeted gravitational-wave searches for compact binary mergers using the GstLAL inspiral pipeline. By incorporating sky localization and timing information from external electromagnetic triggers, we enhance the sensitivity of the search for sub-threshold gravitational-wave signals associated with events such as short gamma-ray bursts. Our approach modifies the standard likelihood ratio ranking statistic to include a sky localization prior, allowing for a more focused analysis on specific regions of the sky. We demonstrate the effectiveness of this method through injection studies, comparing the performance of the targeted search against the standard all-sky search configuration. The results show a significant improvement in detection efficiency for signals consistent with the provided sky location and timing, while maintaining control over false alarm rates. This targeted search framework enables rapid follow-up of electromagnetic transients, facilitating multi-messenger astronomy efforts in the era of advanced gravitational-wave detectors.

\end{abstract}


\acrodef{LSC}[LSC]{LIGO Scientific Collaboration}
\acrodef{LVC}[LVC]{LIGO Scientific and Virgo Collaboration}
\acrodef{LVK}[LVK]{LIGO Scientific, Virgo and KAGRA Collaboration}
\acrodef{aLIGO}{Advanced Laser Interferometer Gravitational-Wave Observatory}
\acrodef{aVirgo}{Advanced Virgo}
\acrodef{LIGO}[LIGO]{Laser Interferometer Gravitational-Wave Observatory}
\acrodef{IFO}[IFO]{interferometer}
\acrodef{LHO}[LHO]{LIGO-Hanford}
\acrodef{LLO}[LLO]{LIGO-Livingston}
\acrodef{O2}[O2]{second observing run}
\acrodef{O1}[O1]{first observing run}
\acrodef{O3}[O3]{third observing run}
\acrodef{O3a}[O3a]{first half of the third observing run}
\acrodef{O3b}[O3b]{second half of the third observing run}
\acrodef{O4a}[O4a]{first half of the fourth observing run}
\acrodef{O4b}[O4b]{second half of the fourth observing run}
\acrodef{O4}[O4]{the fourth observing run}
\acrodef{O5}[O5]{the fifth observing run}

\acrodef{SSM}[SSM]{subsolar-mass}
\acrodef{BH}[BH]{black hole}
\acrodef{BBH}[BBH]{binary black hole}
\acrodef{BNS}[BNS]{binary neutron star}
\acrodef{IMBH}[IMBH]{intermediate-mass black hole}
\acrodef{NS}[NS]{neutron star}
\acrodef{BHNS}[BHNS]{black hole--neutron star binaries}
\acrodef{NSBH}[NSBH]{neutron star--black hole binary}
\acrodef{PBH}[PBH]{primordial black hole binaries}
\acrodef{CBC}[CBC]{compact binary coalescence}
\acrodef{GW}[GW]{gravitational wave}
\acrodef{GWH}[GW]{gravitational-wave}
\acrodef{EM}[EM]{electromagnetic}
\acrodef{DBH}[DBH]{dispasstive black hole binaries}
\acrodef{GRB}[GRB]{gamma-ray burst}
\acrodef{RA}[RA]{right ascension}
\acrodef{DEC}[DEC]{declination}

\acrodef{SNR}[SNR]{signal-to-noise ratio}
\acrodef{LR}[LR]{likelihood ratio}
\acrodef{KDE}[KDE]{kernel density estimate}
\acrodef{FAR}[FAR]{false alarm rate}
\acrodef{VT}[$VT$]{sensitive space-time volume}
\acrodef{PSD}[PSD]{power spectral density}
\acrodef{PDF}[PDF]{probability distribution function}
\acrodef{GR}[GR]{general relativity}
\acrodef{NR}[NR]{numerical relativity}
\acrodef{PN}[PN]{post-Newtonian}
\acrodef{EOB}[EOB]{effective-one-body}
\acrodef{ROM}[ROM]{reduced-order model}
\acrodef{IMR}[IMR]{inspiral--merger--ringdown}
\acrodef{EOS}[EoS]{equation of state}
\acrodef{FF}[FF]{fitting factor}
\acrodef{FT}[FT]{Fourier Transform}

\acrodef{LAL}[LAL]{LIGO Algorithm Library}
\acrodef{GWTC}[GWTC]{Gravitational Wave Transient Catalog}
\acrodef{GDB}[GraceDB]{the Gravitational-wave Candidate Event Database}
\acrodef{FRB}[FRB]{Fast-Radio Burst}
\newcommand{\PN}[0]{\ac{PN}\xspace}
\newcommand{\BBH}[0]{\ac{BBH}\xspace}
\newcommand{\BNS}[0]{\ac{BNS}\xspace}
\newcommand{\BH}[0]{\ac{BH}\xspace}
\newcommand{\NR}[0]{\ac{NR}\xspace}
\newcommand{\GW}[0]{\ac{GW}\xspace}
\newcommand{\SNR}[0]{\ac{SNR}\xspace}
\newcommand{\SSM}[0]{\ac{SSM}\xspace}
\newcommand{\aLIGO}[0]{\ac{aLIGO}\xspace}
\newcommand{\PSD}[0]{\ac{PSD}\xspace}
\newcommand{\GR}[0]{\ac{GR}\xspace}
\newcommand{\EOS}[0]{\ac{EOS}\xspace}
\newcommand{\LVC}[0]{\ac{LVC}\xspace}


\newcommand{\GSTLAL}{GstLAL\xspace}
\newcommand{\IMRPHENOMD}{IMRPhenomD\xspace}
\newcommand{\MANIFOLD}{{\fontfamily{qcr}\selectfont manifold}\xspace}
\newcommand{\SBANK}{{\fontfamily{qcr}\selectfont SBank}\xspace}

\section{Introduction} \label{sec:intro}
The discovery of the \ac{BNS} merger GW170817 by \ac{LIGO} and Virgo detectors together with its \ac{EM}
counterpart marked a milestone for multimessenger astronomy ~\citep{170817_observation, 170817_mm}. This
event demonstrated that joint \ac{GW} and \ac{EM} observations can reveal
detailed information about the physics of extreme phenomena. At the time of writing, the \ac{LVK} network has completed four observing runs and reported 218 detections with astrophysical probability greater than 0.5~\citep{gwtc-5_results}, based on analyses of data up to the second part of \ac{O4}, with two additional data bulks yet to be analyzed. However, among these detections, only one additional \ac{BNS}, i.e., GW190425~\citep{LIGOScientific:2020aai}, has been identified, and no other \ac{EM} counterparts have been observed.

 In preparation for multi-messenger opportunities, the \ac{LVK} has
developed rapid \ac{GW} search pipelines capable of identifying signals from
compact binary coalescences in near real-time, enabling prompt \ac{EM} follow-up
observations~\citep{Abbott_2019}. One of these pipelines, \GSTLAL~\citep{gstlal_cody, gstlal_o2,dtdphi,gstlal_o4_lr,gstlal_o4_offline,gstlal_o4_templatebank}, has
played an integral role in many \ac{LVK} detections to date and powers
low-latency alerts that facilitate dozens of \ac{EM} counterpart searches.
Among these, GW170817~\citep{170817_observation}, GW190425~\citep{LIGOScientific:2020aai}, and GW190814~\citep{LIGOScientific:2020zkf} are particularly notable, as their source classification placed them in regimes where \ac{EM} counterparts are possible, prompting extensive \ac{EM} follow-up campaigns across the spectrum~\citep{gcn_gw170817, gcn_gw190425, gcn_gw190814}.

These standard \ac{GW} searches for compact binaries are typically conducted as
\textit{all-sky} searches, meaning they have no \textit{a priori} knowledge of
where in the sky a signal might originate. The search algorithm must consider
the possibility of a signal from any direction at any time within the observing
period. This all-sky approach is robust for discovering unknown events, but it
comes at the cost of a higher background (many trials) and reduced sensitivity
for any particular location. In contrast, if an external observation of \ac{EM} transients such as a \ac{GRB} or an optical transient provides a sky
localization for a candidate \ac{BNS} merger, one can perform a
\textit{targeted} \ac{GW} search restricted to that sky region and time informed by the \ac{EM} transients. By incorporating
the sky position information from an \ac{EM} counterpart (often provided as a
probability skymap or a specific coordinate), the search can be made more
sensitive to a signal from that location and potentially lower the false alarm
rate by ignoring inconsistent triggers from the background.

Several past works have explored triggered or targeted \ac{GW} searches
coincident with external events, such as \acp{GRB} and \acp{FRB}~\citep{grb_o1, grb_o2, grb_o3a, grb_o3b, frb_o3a}. These analyses typically fix the time of the \ac{GW} search to
a window around the \ac{EM} trigger and assume the sky position in analyzing data from \ac{GW} detectors. In particular, PyGRB~\citep{pygrb1, pygrb2} is a
targeted \ac{GW} search pipeline designed to identify signals associated with
\ac{EM} triggers, in particular short \acp{GRB} and \acp{FRB}, leveraging external information such as trigger
time and sky location. By fixing the coalescence time and direction, PyGRB
performs a \textit{coherent} matched-filtering analysis across multiple detectors,
marginalizing over unknown extrinsic parameters to compute the optimal detection
statistics for a given sky location and time. This targeted approach yields higher sensitivity than all-sky
searches for signals coincident with known sky location. However, the computation of a different detection statistics and its background collection requires re-filtering data for every target, which makes the PyGRB analysis computationally expensive, and hence, intended for offline follow-up.

Furthermore, in the advanced detector era, the paucity of \ac{GW} detections with confirmed electromagnetic counterparts provides strong motivation to configure targeted \ac{GW} searches that can deliver trigger information to external observatories with low to moderate latency. Due to their computational demands and latency, conventional follow-up analyses must impose stringent selection criteria on \ac{EM} triggers, whereas lower-cost searches can afford to follow up a substantially broader set of external triggers.

Here we present a targeted \GSTLAL search for \acp{CBC} that incorporates external \ac{EM} information while retaining a 
low computational cost. Our method modifies the standard \GSTLAL ranking statistic in two ways: by replacing the all-sky coherence PDF with a sky-localized coherence PDF informed by the EM position, and by applying a trigger-time prior centered on the EM trigger time. Because these modifications are implemented within the rerank workflow discussed in~\refref{gstlal_rerank}, the search can be performed without repeating the matched-filtering stage, substantially reducing computational cost relative to conventional targeted follow-up analyses.

The paper is structured as follows: In \secref{sec:method}, we outline the \GSTLAL search workflow and describe our modifications to the ranking statistic, including the incorporation of sky localization information through a targeted coherence \ac{PDF} and a temporal prior based on \ac{EM} trigger times. \secref{sec:results} presents results from injection studies evaluating the performance of the targeted search compared to the standard all-sky configuration, including sensitivity improvements and computational efficiency gains. In \secref{sec:discussion}, we discuss the implications of our findings and outline the necessary steps for deploying this methodology in future observing runs, along with potential extensions for broader applications. Finally, we conclude with a summary of our results and future prospects in \secref{sec:conclusion}.

\section{Method} \label{sec:method}
\subsection{Overview of the \GSTLAL search pipeline}\label{sec:method_gstlal}
\GSTLAL is a \ac{GW} detection pipeline that employs matched filtering to
identify signals from \acp{CBC} in strain data from
multiple detectors~\citep{gstlal_cody, gstlal_o2,dtdphi,gstlal_o4_lr,gstlal_o4_offline,gstlal_o4_templatebank}. The pipeline is designed for low-latency
operation, connecting the GStreamer streaming framework with \ac{LAL} routines, and can also be run in an offline mode for deeper
analysis. The core workflow of \GSTLAL involves several key steps.

First, the pipeline conditions incoming data through calibration and quality
checks, estimates the \acl{PSD} of detector noise, and
applies matched filtering using a pre-computed template bank. To reduce
computational cost, GstLAL employs singular value decomposition to
compress similar waveform templates into efficient orthogonal bases, and multi-banding to sample waveforms efficiently in time-domain, enabling
real-time filtering across a wide parameter space.

When the matched-filter output exceeds a predefined triggering \ac{SNR} threshold, the pipeline records the outputs and the corresponding template parameters as a trigger. The pipeline then searches for coincident triggers across detectors that are consistent in both arrival time and template morphology, forming \ac{GW} candidates. Candidates may also be formed from single-detector triggers in the absence of coincident triggers, provided the single-detector \ac{SNR} exceeds the network \ac{SNR} threshold in addition to the trigger threshold. The identified
candidates are then ranked in terms of a multi-detector ranking statistic,
formulated as a \ac{LR} \citep[see \secref{sec:method_lr} or ][for more details]{O4_lr}. Finally, statistical significance is assessed using simulated trigger
samples randomly drawn from the noise model, allowing the estimation of false alarm rates.

A key feature of \GSTLAL is its \textit{rerank}
workflow~\citep{gstlal_rerank}, which allows the pipeline to decouple the
computationally expensive matched-filtering stage from the ranking and
significance estimation stages. In this modular framework, candidate triggers
identified during the initial filtering step are stored and can be subsequently
reanalyzed with updated or alternative ranking statistics without needing to
repeat the matched-filtering process. This modularity enables rapid testing of
new likelihood models, coherence tests, and prior assumptions such as the
sky-localization priors described in this work, while reusing the same set of
triggers from the filtering stage. The rerank workflow thus provides both
computational efficiency and analytical flexibility, making it particularly
well-suited for targeted searches where external information (e.g., from \ac{EM}
triggers) becomes available after the initial detection pipeline has already
processed the data.

\subsection{Ranking Statistic} \label{sec:method_lr}
As mentioned above, for every trigger the GstLAL pipeline computes a likelihood
ratio to rank candidates from most to least likely to represent real \ac{GW} signals.
This statistic quantifies the probability of obtaining the observed data
($\vec{d}$) under the signal hypothesis ($\mathcal{H}_\mathrm{s}$) relative to
the noise hypothesis ($\mathcal{H}_\mathrm{n}$),
\begin{align}
    \label{eq:lr}
    \mathcal{L}=\frac{P\left(\vec{d}\mid \mathcal{H}_\mathrm{s}\right)}{P\left(\vec{d}\mid \mathcal{H}_\mathrm{n}\right)},
\end{align}
where $\vec{d}$ represents a collection of
various observables recorded for \ac{GW} candidates, such as \ac{SNR},
$\xi^2$-signal-based-veto parameter
\citep{gstlal_cody, O4_lr}. In particular, the signal likelihood in the numerator is
factorized into several components of conditional \acp{PDF} that reflect the
physical and statistical properties expected for true \ac{GW} signals.
Of these components, the likelihood includes a prior \ac{PDF} on the trigger time, $t_\mathrm{ref}$, and a multi-detector joint \ac{PDF}
over matched-filter outputs such as \acp{SNR} observed at each detector ($\vec{\rho}$), coalescence time delays ($\vec{\Delta t}$) and phase differences
($\vec{\Delta \phi}$) across multiple detectors
\begin{align}
    \label{eq:signal_model}
        P(\vec{d} \mid \mathcal{H}_\mathrm{s}) \propto P\left(t_\mathrm{ref} | \mathcal{H}_\mathrm{s}\right) P\left(\vec{\rho}, \vec{\Delta t}, \vec{\Delta \phi} \mid \vec{O}, t_\mathrm{ref}, \mathcal{H}_\mathrm{s}\right),
\end{align}
where $\vec{O}$ is a subset of the $N$ detectors observing an event in coincidence.

A crucial part of this framework is the treatment of $\vec{\Delta t}$ and $\Delta
\phi$ as inter-detector \textit{coherence} tests. For a true astrophysical
signal, the relative arrival times and phases between detectors are tightly
constrained by the source's sky position and polarization.  GstLAL models the
joint \ac{PDF} of $\vec{\Delta t}$ and $\vec{\Delta \phi}$, which we call ``coherence
\ac{PDF}'' hereafter, based on simulated signals, distributed uniformly over all-sky, capturing the characteristic
correlation of these parameters across the network. This distribution forms the
basis of a powerful signal-vs-noise discriminator: noise coincidences tend to
exhibit incoherent or inconsistent timing and phase relationships, whereas real
signals follow closely to the predicted correlations. This coherence test has
been demonstrated to improve the search sensitivity for triggers with the false
alarm probability above $10^{-3}$~\citep{dtdphi}.
As described below, in the all-sky search, this term marginalizes over all sky locations, while our approach in the targeted search is to condition it on the sky location provided by the \ac{EM} trigger. Consequently, candidates are rewarded when their inter-detector timing and phase differences are consistent with the targeted direction and down-ranked when they are not.

\subsection{Sky localization prior} \label{sec:method_dtdphi}
For a source at a known sky location, the relative arrival times ($\vec{\Delta t}$) and phases ($\vec{\Delta \phi}$) measured across detectors are constrained by the detector geometry and the source orientation. In the targeted search, we use this information by replacing the all-sky coherence PDF with the one tailored to the EM-inferred time and source location. Candidate events are therefore also ranked according to how well their observed inter-detector timing and phase differences match the targeted sky geometry.

In the \GSTLAL likelihood formalism, the coherence term can be factorized as
\begin{align}
    \label{eq:dtdphi}
    \begin{aligned}
    P(\vec{\rho}, \vec{\Delta t}, \vec{\Delta \phi} \mid  \vec{O}, t_\mathrm{ref}, \mathcal{H}_\mathrm{s}) &\propto
|\boldsymbol{\mathcal{J}}(\vec{\rho})|\times \rho_\mathrm{net}^{-4}\\
    &\quad\times P(\Delta\vec{\ln \mathcal{D}}, \vec{\Delta t}, \vec{\Delta \phi}\mid ...),
    \end{aligned}
\end{align}
where $\Delta\vec{\ln \mathcal{D}}$ is a $N-1$ dimensional vector of logarithmic
effective distances for each detector relative to that of a reference detector,
i.e. $\Delta\vec{\ln \mathcal{D}}=\vec{\ln \mathcal{D}} - \ln
\mathcal{D}_\mathrm{ref}$~\citep{O4_lr}, and the factor of $\rho_\mathrm{net}^{-4}$ ($\rho_\mathrm{net}\equiv\sqrt{\sum\rho_i^2}$) comes from the \ac{PDF} of
$\rho_\mathrm{net}$ assuming isotropic distribution of \ac{GW} sources in the local Universe ~\citep{kipp_lr, Schutz_2011}.
Also, note that $\boldsymbol{\mathcal{J}}(\vec{\rho})$ is a Jacobian matrix to
take into account the conversion of volume elements due to the following
coordinate transformation $\vec{\rho}\rightarrow(\rho_\mathrm{net},\Delta\vec{\ln \mathcal{D}})$.

Furthermore, the joint \ac{PDF} on $(\Delta\vec{\ln \mathcal{D}}, \vec{\Delta t}, \vec{\Delta \phi})$ is formulated as follows:
\begin{align}
    \label{eq:methods_dtdphi_sum}
    \begin{aligned}
        P&(\Delta\vec{\ln \mathcal{D}}, \vec{\Delta t}, \vec{\Delta \phi}\mid ...) \\
        &= \sum_{\hat{\Omega}, \iota, \psi} P(\Delta\vec{\ln \mathcal{D}}, \vec{\Delta t}, \vec{\Delta \phi}\mid \hat{\Omega}, \iota, \psi) P(\hat{\Omega})P(\iota, \psi),
    \end{aligned}
\end{align}
where $\hat{\Omega}, \iota, \psi$ are the sky location, the inclination angle
and the polarization angle of a hypothetical binary system. $P(\Delta\vec{\ln
\mathcal{D}}, \vec{\Delta t}, \vec{\Delta \phi}\mid \hat{\Omega}, \iota, \psi) $
represents the probability density that the true values of $(\hat{\Omega},
\iota, \psi)$ are recovered as the observed $(\Delta\vec{\ln \mathcal{D}},
\vec{\Delta t}, \vec{\Delta \phi})$ due to noise fluctuation. This \ac{PDF} is
modeled as a multivariate Gaussian distribution, whose covariance matrix is
described in \appref{app:pdf_width}. More crucially,
$P(\hat{\Omega})$ and $P(\iota, \psi)$ are the prior \acp{PDF} of those
parameters respectively. In the standard all-sky configuration, the pipeline
assumes no prior preference for any sky direction, i.e., $P(\hat{\Omega})$ being
isotropic over the sky, and $P(\iota, \psi)$ is assumed to be uniform in
$\cos\iota \in[-1, 1]$ and $\psi\in[0, 2\pi]$. The idea of our targeted search
is to inform the pipeline of the sky location given by an external \ac{EM}
trigger, by providing a localized distribution consistent with the trigger, and
construct the coherence \ac{PDF} based on \eqref{eq:methods_dtdphi_sum}.

\begin{figure*}[htbp]
    \centering
    \begin{minipage}[t]{.8\columnwidth}
        \centering
        \includegraphics[width=\linewidth]{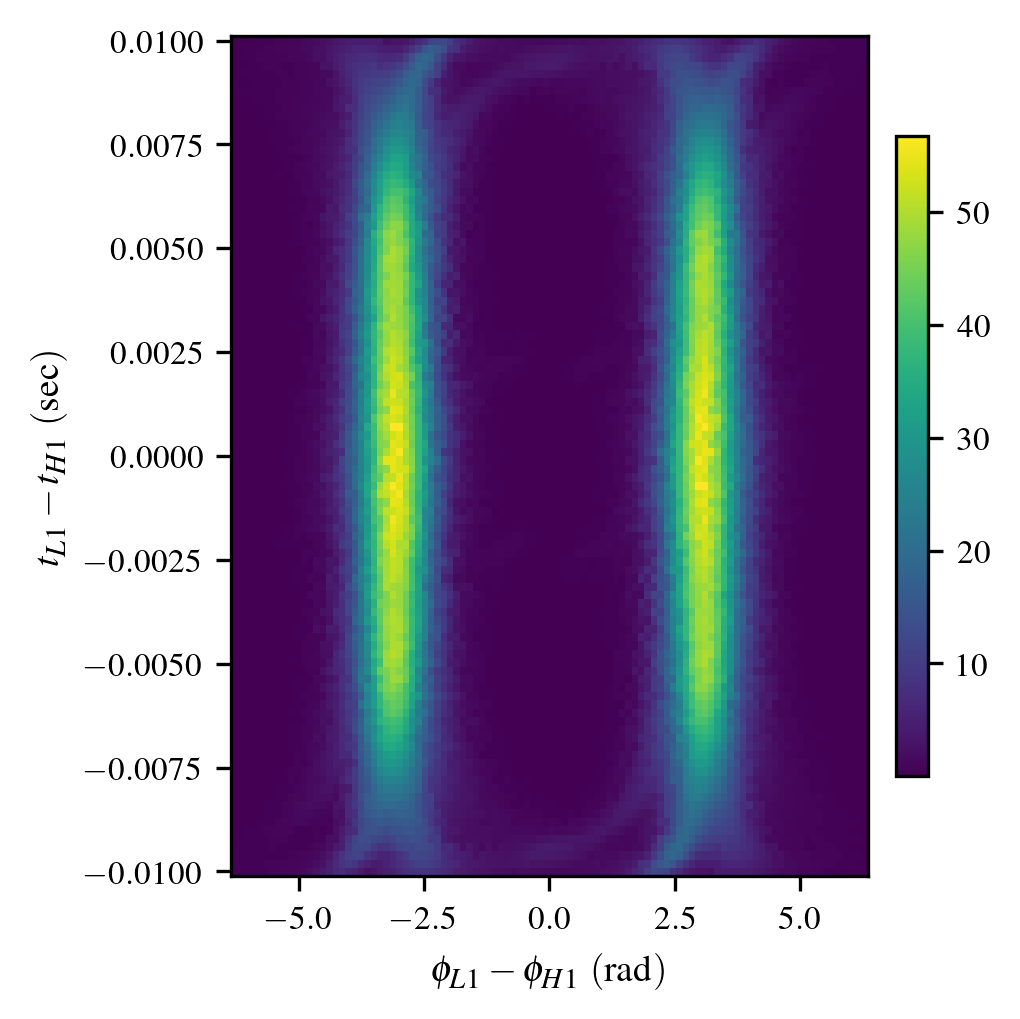}
    \end{minipage}
    \begin{minipage}[t]{.8\columnwidth}
        \centering
        \includegraphics[width=\linewidth]{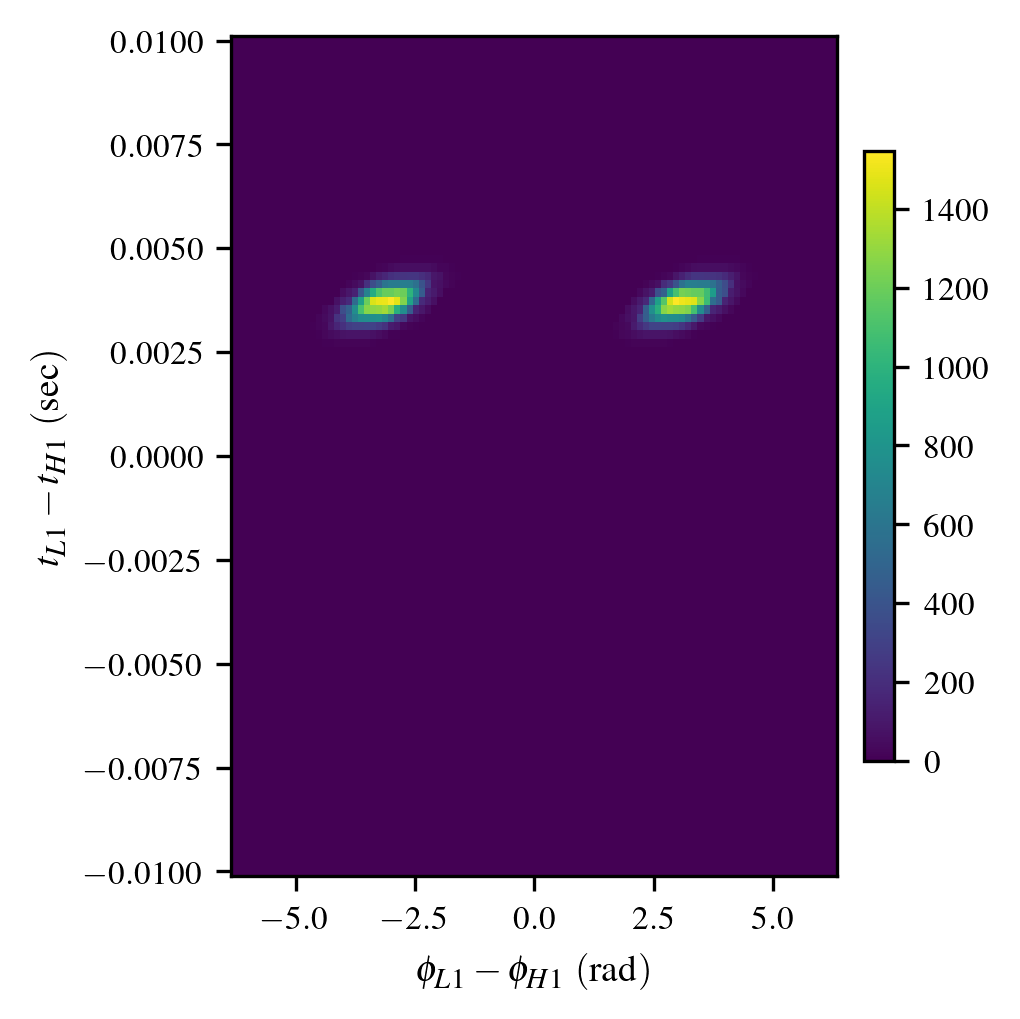}
    \end{minipage}
    \caption{\label{fig:dtdphi_compare}Coherence \ac{PDF} $P(\Delta\vec{\ln \mathcal{D}}, \vec{\Delta t}, \vec{\Delta \phi}\mid ...)$ for the LIGO Hanford (H1) and Livingston (L1) detector pair, shown in the $(\vec{\Delta t}, \vec{\Delta \phi})$ space at a fixed value of $\Delta\vec{\ln \mathcal{D}}\approx 0.15$. Left: All-sky coherence \ac{PDF} used in the standard search, computed by marginalizing uniformly over all sky locations. Right: Single-pixel coherence \ac{PDF}, computed for a specific sky location ($P(\hat{\Omega})=\delta(\hat{\Omega}-\hat{\Omega}^\prime)$, corresponding to one \texttt{healpix} pixel with nside$=8$). The bright spots in the single-pixel \ac{PDF} correspond to the physically allowed combinations of time delay and phase difference for that particular sky location, and these features are present but diluted in the all-sky \ac{PDF} due to marginalization over all directions. This demonstrates the enhanced discriminating power of the targeted search, which can effectively distinguish signals from the target sky location from both noise triggers and \ac{GW} signals originating from other directions. Color intensity represents the normalized probability density.}
\end{figure*}

To demonstrate the effect of the prior \ac{PDF} $P(\hat{\Omega})$, \figref{fig:dtdphi_compare} shows $P(\Delta\vec{\ln \mathcal{D}}, \vec{\Delta t}, \vec{\Delta \phi}\mid ...)$ for LIGO-Hanford (H1) and Livingston (L1) pair, being sliced at $\Delta\vec{\ln \mathcal{D}}\approx 0.15$ to get a two-dimensional \ac{PDF} on $(\vec{\Delta t}, \vec{\Delta \phi})$.\footnote{$\Delta\vec{\ln \mathcal{D}}\approx 0.15$ is motivated by realistic values for the observed \acp{SNR} and horizon distance for the LIGO detector pair for \ac{O4}, e.g., $(\rho_\mathrm{H1}, \rho_\mathrm{L1})=(8, 8)$ and the \ac{BNS} range of 110 Mpc and 140 Mpc for Hanford and Livingston, respectively.} The left plot corresponds to the one used for the standard all-sky search, where $P(\hat{\Omega})$ is isotropic over the sky, while the right plot is derived from only one pixel in the \texttt{healpix} pixelization with nside$=8$, i.e., $P(\hat{\Omega})=\delta(\hat{\Omega}-\hat{\Omega}^\prime)$. Since the all-sky coherence \ac{PDF} is a result of uniform marginalization across all pixels, the bright spots shown in the one-pixel version of the \ac{PDF} are part of its all-sky counterpart, which is associated with the specific time delay between the two detectors given a pixel location. The much narrower structure of the one-pixel \ac{PDF} illustrates a better capability to distinguish \ac{GW} signals originating from the target sky location from not only noise triggers but also other \ac{GW} signals, whose $(\vec{\Delta t}, \vec{\Delta \phi})$ estimates fall out of the bright spots. Furthermore, this coherence test is performed for every detector pair in the network, and hence, additional detectors would provide more coherence information, making the targeted search even more sensitive.

We also note that while we focus on point-like sky localizations (e.g., a \ac{EM} trigger identified in a specific galaxy or coordinates), the method could be extended to broader localizations by integrating $p(\vec{\Delta t}, \vec{\Delta \phi})$ over a \ac{PDF} of sky locations, essentially summing contributions from multiple nearby sky pixels weighted by their probability in the skymap, which we leave for future work.

\subsection{Trigger time prior} \label{sec:method_time_prior}
In addition to incorporating sky localization information, our targeted search framework can leverage the timing information from \ac{EM} triggers to further improve sensitivity and significance estimation. As shown in \eqref{eq:signal_model}, the signal likelihood includes a prior \ac{PDF} on the trigger time, $P(t_\mathrm{ref} | \mathcal{H}_\mathrm{s})$\footnote{Technically, \refref{O4_lr} denotes this \ac{PDF} as $P(t_\mathrm{ref}, \phi_\mathrm{ref} |\theta, \mathcal{H}_\mathrm{s} )$ for completeness. Nevertheless, in practice it is uniform in the reference phase $\phi_\mathrm{ref}$, and we also omit the dependence on the template parameters, $\theta$, which is encoded in the horizon distance $D_H(\theta)$ as it is not relevant in the targeted searches described here.}. In the standard all-sky configuration, this prior is uniform over the analysis time window, reflecting no \textit{a priori} knowledge of when a \ac{GW} signal might occur. However, when an \ac{EM} trigger provides a specific detection time ($t_\mathrm{EM}$), we can incorporate this temporal information by adopting a Gaussian prior centered on the \ac{EM} trigger time:
\begin{align}
    \label{eq:time_prior}
    P(t_\mathrm{ref} \mid t_\mathrm{EM}, \mathcal{H}_\mathrm{s}) \sim \mathcal{N}(\mu = t_\mathrm{EM}, \sigma),
\end{align}
where $\sigma$ characterizes the uncertainty or systematic offset in the
temporal coincidence between the \ac{GW} and \ac{EM} signals.  We substitute
\eqref{eq:time_prior} into \eqref{eq:signal_model} for the full \ac{LR}
calculation, which is now conditioned by the \ac{EM} trigger time,
$t_\mathrm{EM}$. Essentially, this temporal prior down-weights \ac{GW}
candidates that are not coincident with the \ac{EM} trigger, thereby effectively suppressing candidates that occur far from the expected coincidence window.

In theory, the choice of $\sigma$ depends on the physical scenario under
consideration and the characteristics of the \ac{EM} trigger. For example, in
GW170817, the \ac{GW} signal preceded the associated short \ac{GRB} by
approximately 1.7 seconds~\citep{170817_observation}. While a shorter time
window provides better capability to suppress background triggers and
potentially improves search sensitivity, the \GSTLAL analysis requires a
sufficient number of foreground triggers to model the clustering effect
\citep[see Section~III G of][]{gstlal_o4_offline}, which requires $\sigma$ to be
sufficiently large. Given this consideration, we conservatively adopt a Gaussian
prior with $\sigma \sim$ 3600 seconds centered on the \ac{EM} trigger time.
In other words, this width is chosen to satisfy the practical requirements of the
targeted analysis rather than astrophysical constraints, potentially at the cost
of sensitivity. Further optimization of $\sigma$ warrants additional
investigation, which we defer to future work.

\subsection{Implementation} \label{sec:method_implementation}
In general, it is impractical to compute the coherence \ac{PDF} of \eqref{eq:methods_dtdphi_sum} on the fly, which poses a critical challenge for low-latency detection or quick \ac{GW} follow-up. \refref{dtdphi} has formulated an approximate method to precompute the summation with regard to the extrinsic parameters, $(\hat{\Omega}, \iota, \psi)$ shown in \eqref{eq:methods_dtdphi_sum}, and to quickly evaluate the \ac{PDF} given the observed $(\Delta\vec{\ln \mathcal{D}}, \vec{\Delta t}, \vec{\Delta \phi})$, which allows us to implement the coherence test to the \GSTLAL all-sky search in low latency. Utilizing this formalism, in this work, we precompute a \textit{pixel-based} \ac{PDF}, e.g., the right panel of \figref{fig:dtdphi_compare}, for every pixel in the sky and load them from disk upon a targeted search.

The coherence \ac{PDF} depends on the sky location of the \ac{GW} event relative to the detector network. For a fixed astrophysical or \ac{EM} source, this relative location evolves with time due to the Earth's rotation. We precompute the pixel-based coherence \acp{PDF} in the reference frame fixed to Earth. During the reranking procedure for the targeted search, the appropriate pixel \ac{PDF} is then selected based on the \ac{EM} sky location and the actual \ac{GW} trigger time:
\begin{align}
    \label{eq:methods_dtdphi_targeted}
    \begin{aligned}
        P&(\Delta\vec{\ln \mathcal{D}}, \vec{\Delta t}, \vec{\Delta \phi}\mid t_\mathrm{ref}, \hat{\Omega}_0, ...) \\
        &= \sum_{\iota, \psi} P\left(\Delta\vec{\ln \mathcal{D}}, \vec{\Delta t}, \vec{\Delta \phi}\mid \hat{\Omega}(t_\mathrm{
        ref}), \iota, \psi\right)P(\iota, \psi),
    \end{aligned}
\end{align}
where $\hat{\Omega}_0 = (\alpha_0, \delta_0)$ is the celestial coordinate of a given \ac{EM} trigger and
\begin{align}
    \label{eq:pixel_translation}
    \hat{\Omega}(t_\mathrm{GW}) = \left(\alpha_0+ 360^\circ\frac{t_\mathrm{GW}}{1 \mathrm{\ sidereal\ day}}, \delta_0 \right)
\end{align}
is the apparent location of the GW trigger at $t=t_\mathrm{GW}$, translated from the celestial coordinates of the \ac{EM} trigger.

While setting up a targeted search, we identify a subset of pixels with the same
declination as the \ac{EM} sky location ($\delta_0$) and let the analysis load them into
memory. Following \eqref{eq:pixel_translation}, the analysis constructs an
internal mapping between the \ac{GW} trigger time and the apparent location of the
pixel, and hence, a specific pixel-based \ac{PDF}. Therefore, during the \ac{LR}
evaluation, the analysis can identify the correct \ac{PDF} to use for each candidate according to its
trigger time so that the ranking statistics are optimized for the targeted
pixel. This is done for all candidates in the observational period.
Additionally, as described in \secref{sec:method_time_prior}, the provided
\ac{EM} trigger time determines a functional form of the trigger-time prior in
\eqref{eq:time_prior}, assigning a Gaussian temporal weight to each \ac{GW}
trigger based on its offset from $t_\mathrm{EM}$. We also note that this \ac{LR}
formalism, including both the sky localization and trigger-time priors, is
applied to both foreground and background triggers, so our background estimation
and foreground evaluation are performed consistently, ensuring that the
background estimation properly accounts for the reduced trials factor.  By
combining both these two priors, the targeted search can account for
spatiotemporal coincidence, maximizing the gain in detection efficiency for
genuine multimessenger events while maintaining robustness against false
associations.

\section{Results} \label{sec:results}
The performance of the targeted search is evaluated via two different setups: one where all simulated signals are located at the same sky location and the other where signals are distributed uniformly in the sky. 
\subsection{Localized injection campaign}
\subsubsection{Simulation setup} \label{sec:results_locinj_setup}
For the localized-signal study, we consider six distinct configurations, each corresponding to a fixed sky location defined in the celestial system. We used a 2-day data set from the \ac{O3}, 2019 April 19 16:39 UTC
up to 2019 April 21 16:39 UTC.
We simulate \ac{GW} signals from \ac{BNS} systems (hereafter referred to as \textit{injections}) and add them to the \ac{O3} data. For each injection set corresponding to a fixed sky location, we perform an offline rerank analysis using the coherence \ac{PDF} for both the all-sky and targeted configurations.

We compare the injection recovery and the detection efficiency between the
all-sky and targeted configurations for all six sky locations. To enable a
statistical assessment of injection recovery performance without requiring
multiple rank analyses tied to different \ac{EM} trigger times, we disable the
trigger-time prior described in \secref{sec:method_time_prior}.  Therefore, the
performance improvement shown here is driven solely from the effect of the sky
localization prior.

We simulate 9457 \ac{BNS} injections using the SpinTaylorT4 approximant \citep{buonanno2009comparison}. For matched filtering, two waveform models are employed depending on the system's chirp mass: the frequency-domain, post-Newtonian TaylorF2 model \citep{buonanno2009comparison} for chirp masses between $0$ and $1.73~M_\odot$, and the frequency-domain reduced-order SEOBNRv4ROM model \citep{bohe2017improved} for chirp masses above $1.73~M_\odot$. The component \ac{NS} masses for the injections are drawn uniformly between $1~M_\odot$ and $3~M_\odot$, with a maximum total mass of $6~M_\odot$. The dimensionless \ac{NS} spins are uniformly distributed between $0$ and $0.05$. The injected signals are distributed uniformly in redshift up to a maximum of $0.15$ (corresponding to a luminosity distance of $\sim 700~\mathrm{Mpc}$).

Among the six sky locations considered, we focus here on the location of GW170817; results for the remaining cases are presented in \appref{app:multi_locations}. In this configuration, all injections share the same celestial coordinates, with \ac{RA} $= 13^{\rm h},9^{\rm m},43.3^{\rm s}$ and \ac{DEC} $= -23^{\circ},22',58''$ \citep{GW170817_MM}, up to small perturbations. As discussed earlier, although the injections are fixed in celestial coordinates, their apparent positions in the Earth-fixed frame vary due to the Earth's rotation. We therefore compute the corresponding apparent sky location for each injection using \eqref{eq:pixel_translation} based on its trigger time, and select the appropriate coherence PDF for ranking. Since coherence PDFs for all sky pixels are precomputed, this procedure incurs only minimal bookkeeping, mapping trigger times to the corresponding PDFs, without significant computational overhead during the ranking stage.

\subsubsection{Injection recovery}
\label{sec:results_locinj_rec}
We first compare the number of injections recovered with a \ac{FAR} below one per month, which represents the high-significance threshold for public alerts after accounting for trials. The all-sky search recovered 818 out of 9{,}457 injections, while the targeted search recovered 859, representing an improvement of approximately 5\%. Note that the majority of injected signals were missed simply because they were simulated at luminosity distances too large to be detectable. These results are consistent with our theoretical expectation that the targeted coherence \ac{PDF} improves injection recovery relative to the all-sky search.
We have also conducted preliminary tests indicating that the width of the targeted coherence \ac{PDF} plays a crucial role in injection recovery and must be tuned carefully to optimize performance. For example, the all-sky coherence \ac{PDF} (see Fig.~\ref{fig:dtdphi_compare}) has a narrow width (computed using \ac{SNR} thresholds of $5, 7$, and $4$ for LIGO Hanford, LIGO Livingston, and Virgo respectively), whereas the coherence \ac{PDF} used in this localized injection campaign has a broader width (computed using \ac{SNR} thresholds of $1.25, 1.75$, and $1$ for LIGO Hanford, LIGO Livingston, and Virgo respectively) to improve recovery. A more detailed assessment is provided in \appref{app:pdf_width}.

\begin{figure}[htbp]
    \centering
    \includegraphics[width=\linewidth]{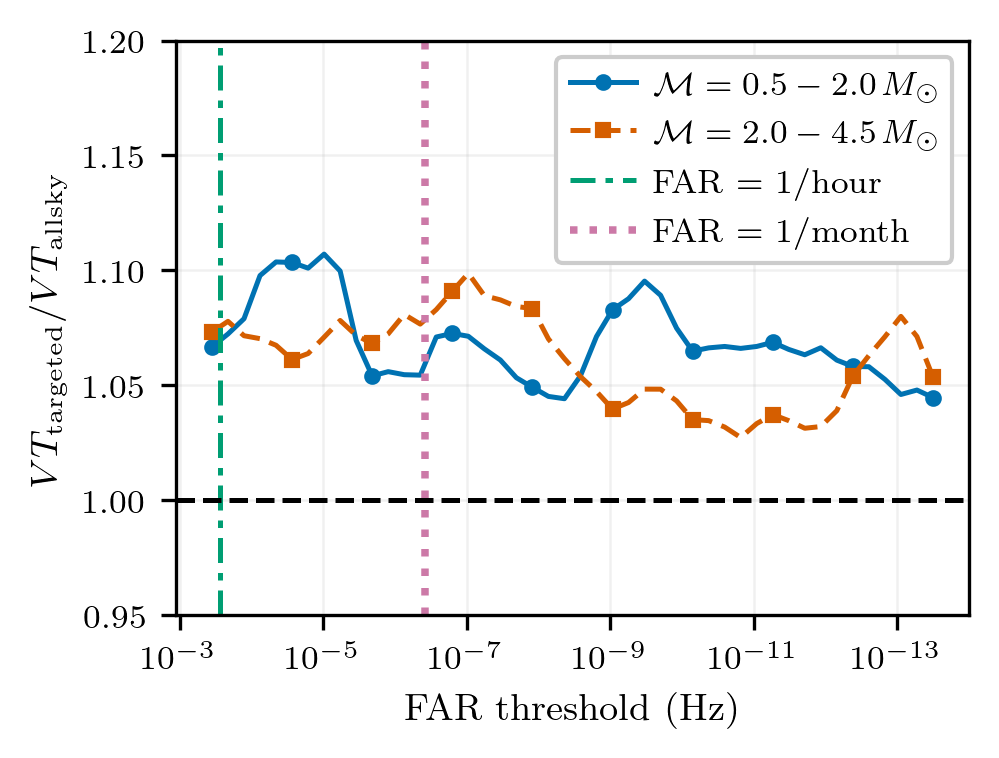}
    \caption{\label{fig:vt_ratio} Ratio of \ac{VT} between the targeted and all-sky searches as a function of \ac{FAR} thresholds. The blue and orange curves represent injections with chirp masses ($\mathcal{M}$) in the ranges $0.5$--$2.0~M_\odot$ and $2.0$--$4.5~M_\odot$, respectively. The horizontal dashed black line at unity indicates equal performance between the two search configurations. Values above unity demonstrate improved \ac{VT} for the targeted search. The vertical dashed red line marks a \ac{FAR} threshold of 1/hour ($\sim 2.8\times10^{-4}~\mathrm{Hz}$), which is the typical threshold for uploading candidates to \ac{GDB}. The vertical dashed purple line marks the FAR threshold of 1/month ($\sim 3.8\times10^{-7}~\mathrm{Hz}$), which is the high significance threshold of \ac{GDB}.}
\end{figure}

Apart from comparing the raw counts of recovered injections, we also estimate the
\ac{VT} of the search in both the all-sky and targeted configurations. The \ac{VT} is a measure that combines the surveyed volume of space and the observation time representative of the number of events that the pipeline can effectively detect above a given threshold, defined as
\begin{align}
    \label{eq:vt}
    VT(\mathrm{FAR}) = T\int_{0}^{\infty}\epsilon(z, \mathrm{FAR})\frac{d V_c(z)}{dz}\frac{1}{1+z}\mathrm{d} z,
\end{align}
where $T$ is the duration of a simulated observation, $\epsilon(z,
\mathrm{FAR})$ is the detection efficiency for the \ac{GW} signals
injected at the redshift in [$z, z+\mathrm{d} z$] and recovered at \ac{FAR}
below a given threshold, and $V_c(z)$ is the comoving volume at the redshift of
$z$.
\figref{fig:vt_ratio} shows the \ac{VT} for the targeted search relative to the
one for the all-sky search as a function of \ac{FAR} thresholds for two chirp-mass ($\mathcal{M}$) ranges. We find that at the
\ac{FAR} threshold of one per month ($\sim10^{-7}~\mathrm{Hz}$), the $\ac{VT}$ improves by
around 8 to 10\%, which is consistent with the raw recovery counts. This improvement effectively extends the search horizon distance, allowing the pipeline to recover a greater number of sub-threshold triggers that would otherwise remain undetected.

\begin{figure}[htbp]
    \centering
    \includegraphics[width=\linewidth]{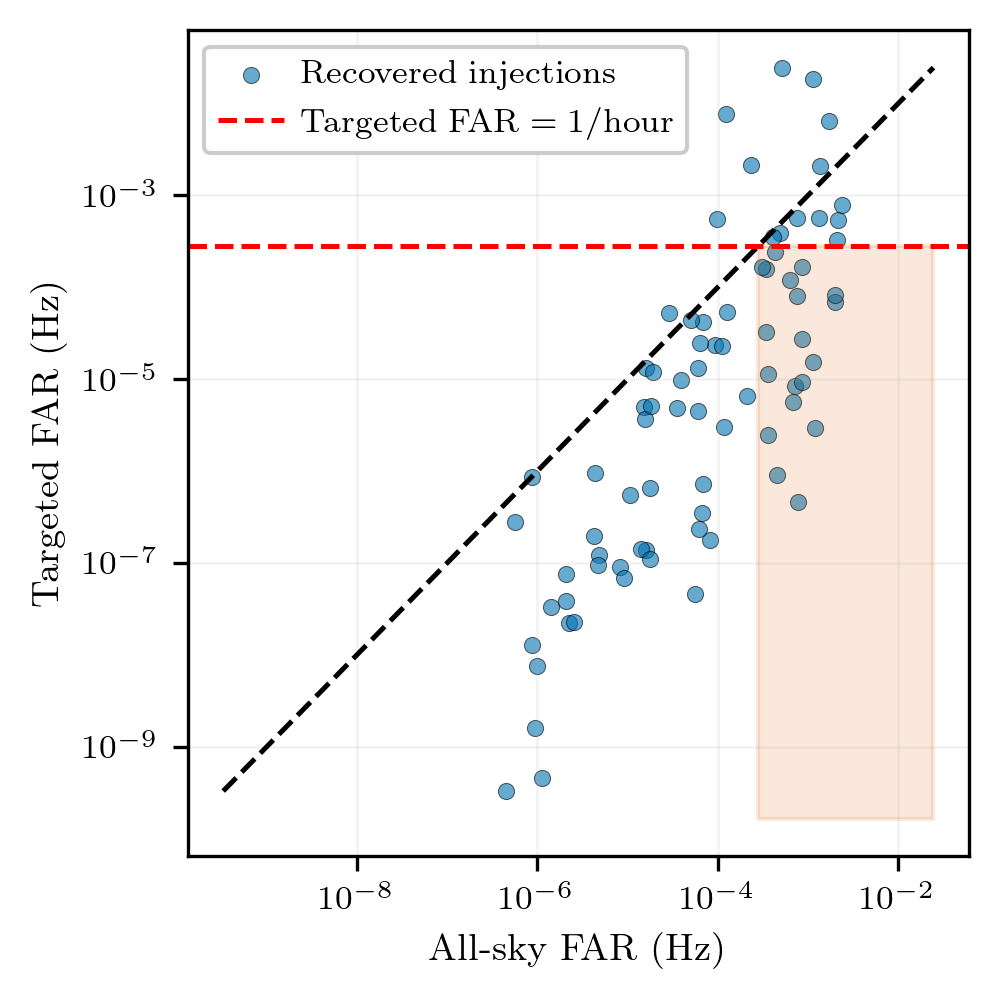}
    \caption{\label{fig:scatter_far}Comparison of \ac{FAR} for recovered injections between the all-sky search ($x$-axis) and the targeted search ($y$-axis). Each point represents an injection recovered by both searches. The horizontal dashed red line marks a \ac{FAR} threshold of 1/hour ($\sim 2.8\times10^{-4}~\mathrm{Hz}$), which is the typical threshold for uploading candidates to \ac{GDB}. The diagonal black dashed line indicates equal \ac{FAR} in both searches. Points below this line represent injections with improved (lower) \ac{FAR} in the targeted search. A significant number of injections with the all-sky $\mathrm{FAR}> 1/\mathrm{hour}$ achieve the targeted $\mathrm{FAR} < 1/\mathrm{hour}$ as illusrated by the orange box, demonstrating the potential for the targeted search to promote sub-threshold candidates to \ac{GDB}-uploadable significance.}
\end{figure}

Another important consideration is pipeline performance at a \ac{FAR} thresholds relevant for uploading candidates to the Gravitational-Wave Candidate Event Database (\ac{GDB}). The \ac{LVK} currently uploads candidates with a \ac{FAR} below approximately one per hour (corresponding to $\sim
2.8\times10^{-4}~\mathrm{Hz}$) to \ac{GDB} for internal records, and issues public alerts to the astronomical community for candidates with a \ac{FAR} below two per day~\citep{emfollow}. As shown in \figref{fig:scatter_far}, the targeted search yields a modest but consistent improvement in \ac{VT} around these thresholds, with gains of approximately 5--10\% relative to the all-sky search. This enhancement translates directly into an increased number of candidates that would be uploaded to \ac{GDB} when conducting targeted follow-up searches based on \ac{EM} triggers.
\figref{fig:scatter_far} further illustrates this point by comparing the
\ac{FAR} of the recovered injections with \ac{FAR} between 1/month and 10/hour
recovered with the all-sky configuration.  It is evident that a
significant number of injections recovered with \ac{FAR} above 1/hour in the
all-sky search fall below the threshold in the targeted search, e.g., the most
noticeable \ac{FAR} improvement by around three orders of magnitudes.  Regarding
the injections recovered by the targeted search with a \textit{higher} \ac{FAR},
we found that they are mostly associated with the injections with a very low
\ac{SNR} (e.g., network \ac{SNR} below 6), whose time and phase measurement
uncertainties are large, and hence, their point estimates can be far from the
bright spot in the coherence \ac{PDF}, leading to a worse ranking statistic and a
higher \ac{FAR}.  Overall, these results indicate that the targeted search not
only improves the signal ranking but also suppresses noise background because
noise triggers are random in $\vec{\Delta t}$ and $\vec{\Delta \phi}$, and hence tend to be
down-ranked by the targeted coherence \ac{PDF}.

The improved recovery of sub-threshold \ac{GW} candidates at these relevant \ac{FAR} levels has important implications for multimessenger astronomy workflows. In particular, it creates a natural synergy with the RAVEN pipeline~\citep{raven1,raven2}, which performs coincidence analyses between \ac{GW} candidates uploaded to GraceDB and external astronomical triggers. By increasing the recovery rate of \ac{GW} candidates in the presence of \ac{EM} information, the targeted search can strengthen joint detection significance. In addition, by providing a larger pool of sub-threshold candidates for RAVEN to follow up on, it complements existing infrastructure and enables more effective coordination of follow-up observations.

\subsubsection{Computational costs}
A critical advantage of the targeted search approach lies in its computational efficiency through the reranking workflow. The standard \GSTLAL pipeline operates in two distinct stages: the filtering stage, where matched filtering is performed against the full template bank to identify candidate triggers, and the ranking stage, where triggers are evaluated using the likelihood ratio statistic. For a typical 2-day analysis segment presented in this section, the complete filtering workflow requires approximately 2000 CPU hours to process the data through matched filtering, coincidence finding, and initial trigger generation across all templates and detectors.

In contrast, our targeted search methodology leverages the modularity of the \GSTLAL pipeline by employing a reranking-only workflow. Since the matched filtering and trigger identification have already been performed in the initial all-sky search, the targeted search can simply reload the existing triggers and re-evaluate them using the modified coherence \ac{PDF} tailored to the specific sky location. This reranking process requires only approximately 40 CPU hours for the same 2-day segment, realizing a reduction in computational cost by a factor of $\sim$50. This dramatic computational savings makes it feasible to perform targeted searches and rapidly respond to external alerts. The efficiency of this approach is particularly valuable in multimessenger follow-up scenarios where timely \ac{GW} analysis is essential for coordinating observations across the \ac{EM} spectrum.

\subsection{Realistic follow-up study} \label{sec:results_real}
To evaluate the targeted search performance in a more realistic multimessenger
follow-up scenario, we conducted an additional test using 8 days of the \ac{O3}
data from 2019 April 19 16:39 UTC up to 2019 April 21 16:39 UTC. Unlike the
localized injection campaign described in \secref{sec:results_locinj_rec}, this
test employed injections distributed uniformly across the entire sky, mimicking
the actual distribution of astrophysical \ac{BNS} mergers. From this all-sky
injection set, we selected a subset of the injections to serve as simulated
\ac{EM} trigger candidates, representing hypothetical \acp{GRB} or other
\ac{EM} transients that would prompt a targeted \ac{GW} follow-up search.

\begin{figure}[htbp]
    \centering
    \includegraphics[width=\linewidth]{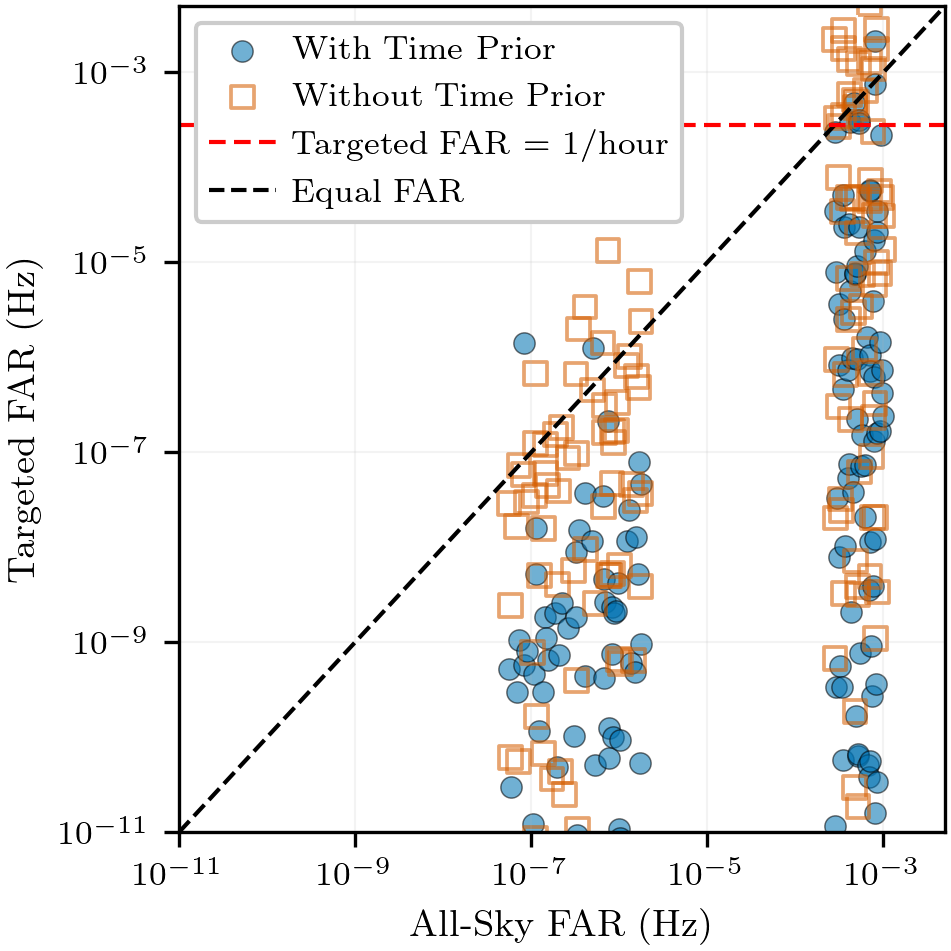}
    \caption{\label{fig:scatter_far_broad_inj}Comparison of recovered injections in the realistic follow-up test, showing targeted-search \ac{FAR} ($y$-axis) versus all-sky \ac{FAR} ($x$-axis). Blue points correspond to targeted reranking with the trigger-time prior enabled, and orange points show reranking without the trigger-time prior. The black dashed diagonal marks equal \ac{FAR} in the two searches, while the red dashed horizontal line indicates the nominal threshold of $1/\mathrm{hour}$ ($\approx 2.8\times10^{-4}\,\mathrm{Hz}$) to upload to \ac{GDB}. Points below the diagonal indicate improved ranking (lower \ac{FAR}) in the targeted search.}
\end{figure}

First, we performed an all-sky search as a control run for the injections, and down-selected 100 injections recovered with a \ac{FAR} around a given threshold. In particular, we focus on events near two representative thresholds: one per month ($\sim 3.8\times10^{-7}\,\mathrm{Hz}$) and one per hour ($\sim 2.8\times10^{-4}\,\mathrm{Hz}$), corresponding to the typical public-alert significance and the internal GraceDB upload threshold, respectively. Second, we followed up each selected injection with a targeted search for using its trigger time and sky location to incorporate the appropriate coherence \ac{PDF}, as described in \secref{sec:method_dtdphi}. This approach simulates the operational workflow where an external \ac{EM} observatory reports a transient event with a known sky position, triggering a rapid targeted \ac{GW} analysis. \figref{fig:scatter_far_broad_inj} shows a comparison of \acp{FAR} between the all-sky and targeted configurations for triggers identified by the three-detector network (\ac{LIGO} Hanford, \ac{LIGO} Livingston, and Virgo). Overall, the targeted searches show comparable or even better \ac{FAR} improvement than the case of localized injection campaign shown in \secref{sec:results_locinj_rec}, where all the injections are ranked within one targeted analysis. The reason why this study with the broad injection set tends to show the \ac{FAR} improvement with tens of order of magnitudes is just because the injection samples we targeted have relatively higher recovered \acp{SNR}, i.e., $\geq10$. While the dramatic \ac{FAR} improvement for some injections simply results from the way we selected the injection samples, the recovered \acp{SNR} of $\sim10$ are still relevant for the realistic follow-up scenario, and hence indicate the strong potential of this targeted search to promote sub-threshold candidates.

\begin{figure}[htbp]
    \centering
    \includegraphics[width=\linewidth]{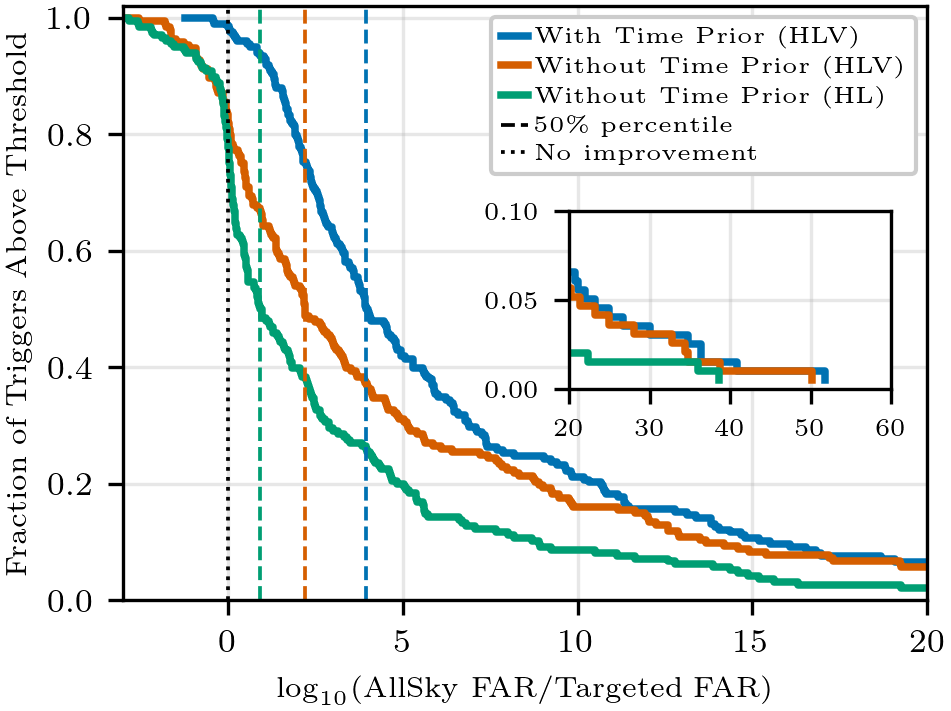}
    \caption{\label{fig:far_improvement_broad_inj} Cumulative distribution of \ac{FAR} improvement for recovered injections in the realistic follow-up scenario as a function of $\log_{10}(\mathrm{FAR}_{\mathrm{allsky}}/\mathrm{FAR}_{\mathrm{targeted}})$. Blue and orange histograms correspond to injections recovered with and without the trigger-time prior using the detector detector network (HLV), respectively, and the green represents the case without the trigger-time prior using the two \ac{LIGO} detectors (HL). While the vertical dotted line at zero marks equal ranking performance between the targeted and all-sky searches, the dashed vertical lines represent 50\% percentile of the distribution for each configuration indicated by the color. The inset shows the strong positive tail above 20 orders-of-magnitude of the \ac{FAR} improvement, demonstrating substantial \ac{FAR} reductions for a subset of events in both configurations.}
\end{figure}

\figsref{fig:scatter_far_broad_inj}{fig:far_improvement_broad_inj} also compares results with and without the trigger-time prior discussed in Section \ref{sec:method_time_prior}, showing that its inclusion yields larger \ac{FAR} improvements. In particular, each dashed vertical line in \figref{fig:far_improvement_broad_inj} denotes the logarithmic \ac{FAR} improvement at 50\% percentile for the configuration indicated by the color, implying the lower bound of \ac{FAR} improvements for half of the recovered injections.
This demonstrates the significant contribution of the trigger-time prior to the overall performance improvement of the targeted search, which is expected because the trigger-time prior can further suppress noise triggers that are not coincident with the \ac{EM} trigger.
Additionally, we show the \ac{FAR} improvement for the triggers recovered by only two \ac{LIGO} detectors (HL) as opposed to the three detectors including Virgo (HLV), suggesting that the triggers found by the three detectors tend to yield even better \ac{FAR} improvements than the two-detector recoveries because the third detector brings more capability to constrain the parameter space in the targeted coherence \ac{PDF} and to distinguish signals from noise.
This underscores the importance of expanded detector network for maximizing the benefits of targeted searches.


\section{Discussion} \label{sec:discussion}
Our results demonstrate the feasibility and benefits of a targeted \ac{GW} search that incorporates \ac{EM}-derived sky localization and timing information within the \GSTLAL pipeline. Here we discuss the implications of these findings and outline directions for future studies and improvements.

Toward the deployment of this targeted search method in a future observing run, several key steps are necessary: First, we plan to conduct this targeted search with real \ac{EM} triggers during \ac{O4} to validate its performance in an operational configuration and potentially detect \ac{GW} counterparts. This will involve archival \ac{EM} trigger database to identify suitable transient events and conducting targeted analyses in response. Second, we will implement this method into a next-generation \ac{GW} search pipeline, SGNL~\citep{sgnl}, which is designed for low-latency operation and GPU acceleration. This migration will be a critical path for this targeted search to be deployed in \ac{O5}. Finally, we will need to establish a robust low-latency infrastructure to receive \ac{EM} trigger alerts, automatically initiate targeted searches, and disseminate results to the astronomical community for follow-up observations. This includes testing the end-to-end workflow from \ac{EM} trigger reception to \ac{GW} candidate generation and alert distribution.

\subsection{Future improvements}
The work presented here also opens several avenues for further investigation and potential improvements to the targeted search methodology. These include optimizing key components of the statistical framework, such as the width of the coherence \ac{PDF} and the trigger-time prior; improving computational efficiency through techniques like background bootstrapping; and extending the framework to more complex astrophysical scenarios, including less well-localized counterparts. 
\begin{itemize}
    \item \textit{Optimizing the width of the coherence \ac{PDF}}\\
    As discussed in \secref{app:pdf_width}, the width of the coherence \ac{PDF} plays a crucial role in balancing sensitivity and robustness. We will explore a range of $\vec{\Delta t}$--$\vec{\Delta \phi}$ prior widths to identify an optimal choice that maximizes detection efficiency while minimizing false alarms. 
    \item \textit{Narrowing time window of the trigger-time prior}\\
    In this study, we adopt a time window of $\sigma \sim 3600$ seconds for the trigger-time prior, which is a conservative choice to ensure we collect sufficient number of foreground triggers for robust modeling of the clustering effect (see \secref{sec:method_time_prior}). While this conservative choice reduces the chance of missing potential signals with larger time offsets due to missing its temporal coincidence, many multimessenger events are expected in practice to exhibit much smaller offsets, on the order of seconds to minutes~\citep{Metzger_2012, Metzger_2019, Nakar_2020}. By narrowing this time window, we can significantly reduce the background of noise triggers that are not temporally coincident with the \ac{EM} event, thereby improving the overall sensitivity of the search. Therefore, we will investigate the optimal width of this time window based on astrophysical models and pipeline performance.
    \item \textit{Background bootstrapping for faster follow up}\\
    While the reranking workflow is computationally efficient, construction of the background distribution for each targeted search, e.g., targeted sky location and trigger time, is still time-consuming, which accounts for more than 90\% of the total CPU hour. To circumvent this bottleneck, we will explore methods to bootstrap the background distribution across different sky locations without computing it from scratch. For instance, we can assess the variation of the background distribution with respect to sky location and determine if a single representative background model can be used for a range of sky locations without significant loss of accuracy. If successful, this approach could eliminate the need to recalculate the background for every follow-up search, potentially enabling near real-time production of targeted search results
    \item \textit{Extensions to less localized sources}\\
    The work presented here focuses on a ``point-source'' counterpart, but the methodology can naturally be extended to scenarios in which the counterpart is not confined to a single sky location. For example, if a pair of \ac{GW} lensed events are suspected, one could perform a targeted search using the predicted sky location of the first \ac{GW} image. Alternatively, if the \ac{EM} localization is provided as an extended region (e.g., a $90\%$ credible contour spanning several tens of square degrees), one could construct a $\vec{\Delta t}$--$\vec{\Delta \phi}$ prior marginalized over that region. One possible approach is to represent the sky localization in a spherical-harmonic basis. Any function on the two-dimensional sphere can be expanded in terms of spherical-harmonic modes and their corresponding coefficients. The coherence \ac{PDF} could then be precomputed for each basis mode, with the appropriate weighted combination constructed on the fly using the coefficients of a given \ac{EM} skymap.

\end{itemize}

\section{Conclusion} \label{sec:conclusion}
We have demonstrated a targeted \ac{GW} search methodology for \ac{BNS} mergers that incorporates external \ac{EM} trigger information within the \GSTLAL pipeline. By modifying the ranking statistic with a sky localization prior and a trigger-time prior, and employing a reranking workflow, the targeted search achieves a $\sim$50-fold reduction in computational cost. Simulation studies using \ac{O3} data showed 8--10\% improvement in \ac{VT} by the targeted search across a range of \ac{FAR} thresholds. Realistic follow-up tests revealed \ac{FAR} improvements spanning decades of orders of magnitude for \ac{SNR} $\geq 10$, which can be improved even further by incorporating the trigger-time prior.

Our investigations found that coherence \ac{PDF} width plays a crucial role: overly restrictive priors degrade performance while appropriately broad priors balance sensitivity with parameter uncertainties. Also, expanded detector networks amplify benefits, with three-detector recoveries showing larger improvements than two-detector cases. We plan to apply this method to archival \ac{O4} \ac{EM} triggers and integrate it into the SGNL~\citep{sgnl} pipeline for \ac{O5} deployment. Future refinements include optimizing the coherence \ac{PDF} width and time window, developing background bootstrapping for faster follow-up, and extending to broader \ac{EM} localizations.

This work establishes a practical pathway for multimessenger astronomy where \ac{EM} observations enhance \ac{GW} search sensitivity. The targeted search with reranking workflow enables detection of fainter signals and faster event confirmation, strengthening joint \ac{GW}-\ac{EM} discovery as global observatory networks expand.

\begingroup
\makeatletter
\let\internallinenumbers\relax
\makeatother
\begin{acknowledgments}
    \AckText{}
    The authors are grateful for computational resources provided by the LIGO
    Laboratory and supported by National Science Foundation Grants PHY-0757058
    and PHY-0823459.  This material is based upon work supported by NSF's LIGO
    Laboratory which is a major facility fully funded by the National Science
    Foundation.  LIGO was constructed by the California Institute of Technology
    and Massachusetts Institute of Technology with funding from the National
    Science Foundation (NSF) and operates under cooperative agreement
    PHY-1764464.This paper
    carries LIGO Document Number LIGO-P2600114.
\end{acknowledgments}
\endgroup

\appendix
\section{Study on the width of the coherence \ac{PDF}}
\label{app:pdf_width}
In this section, we investigate the effect of the coherence \ac{PDF} width on injection recovery. Following \refref{dtdphi}, the probability distribution of the parameters $\vec{\lambda}$ can be expressed as:

\begin{equation}
    \label{eq:parameter_PDF}
    P(\vec{\lambda} \mid \vec{O}, s, \vec{\lambda}_{mi}) = 
    \frac{1}{\sqrt{(2\pi)^3 |\Sigma_{\vec{\lambda}}|}} \exp\left[-\frac{1}{2} (\vec{\Delta\lambda}_i)^T \Sigma_{\vec{\lambda}}^{-1} \vec{\Delta\lambda}_i \right],
\end{equation}
\noindent where $\vec{\lambda} \equiv \{ \Delta \ln D_{\text{eff}}, \vec{\Delta t}, \vec{\Delta \phi} \}$ is the vectorized difference in the time, phase and logarithmic effective distance between a pair of detectors, and $\vec{\Delta\lambda}_i \equiv \vec{\lambda} - \vec{\lambda}_{mi}$ represents the deviation between the actual parameter value and a grid point in parameter space. The width of the PDF is governed by the covariance matrix, $\Sigma_{\vec{\lambda}}$. Assuming the measurements of these observables are independent across detectors, the covariance matrix can be derived from the sum of that of each detector:

\begin{align}
    \left(\Sigma_{\vec{\lambda}}\right)_{ij} = \sigma^2_{\lambda_i\lambda_j} = \sigma^{2(\text{ifo1})}_{\theta_i\theta_j} + \sigma^{2\text{(ifo2)}}_{\theta_i\theta_j}.
\end{align}

We approximate the measurement uncertainty for an individual detector using the inverse of the Fisher information matrix:

\begin{align}
    \Sigma_{\vec{\theta}} &\equiv 
    \begin{bmatrix}
        \sigma^2_{tt} & \sigma^2_{t\phi} & \sigma^2_{t \ln D_{\text{eff}}} \\
        \sigma^2_{\phi t} & \sigma^2_{\phi\phi} & \sigma^2_{\phi \ln D_{\text{eff}}} \\
        \sigma^2_{\ln D_{\text{eff}} t} & \sigma^2_{\ln D_{\text{eff}} \phi} & \sigma^2_{\ln D_{\text{eff}} \ln D_{\text{eff}}}
    \end{bmatrix} 
    = \frac{1}{\rho^2}
    \begin{bmatrix}
        \frac{1}{(2\pi \sigma_f)^2} & \frac{\bar{f}}{2\pi  \sigma_f^2} & 0 \\
        \frac{\bar{f}}{2\pi \sigma_f^2} & \frac{\bar{f}^2}{( \sigma_f)^2} & 0 \\
        0 & 0 & 1
    \end{bmatrix},
\end{align}

where $\sigma_f$ is the effective bandwidth of the signal given by

\begin{align}
    \sigma_f^2 \equiv \bar{f^2} - (\bar{f^1})^2
\end{align}

using the frequency moments of the signal

\begin{align}
    \bar{f^n} \equiv 4\int^\infty_0 df\frac{|\tilde{h}(f)|^2}{S(f)}f^n.
\end{align}

\begin{figure}[h]
    \centering
    \includegraphics[width=0.9\linewidth]{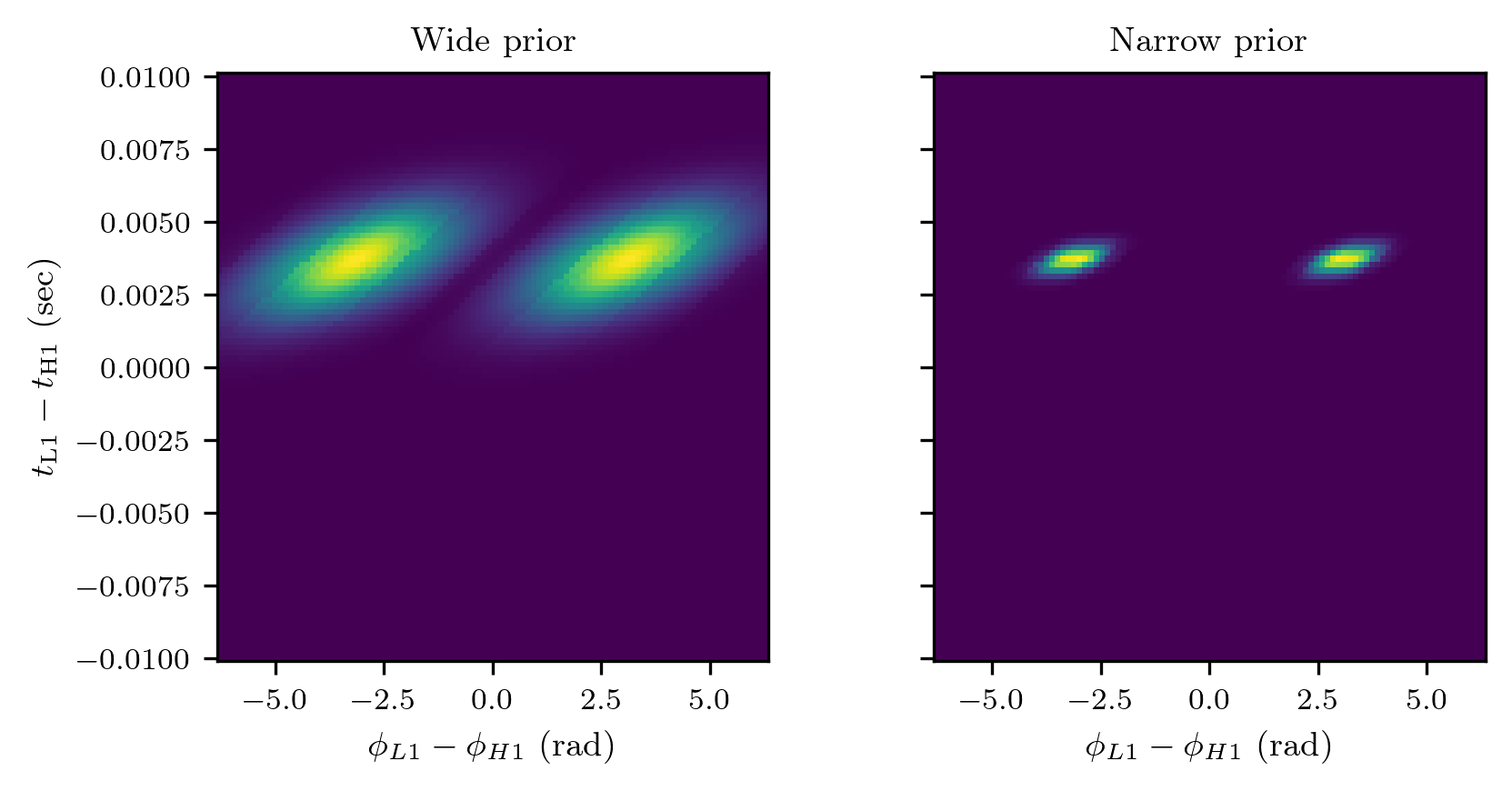}
    \caption{\label{fig:dtdphi_compare_width} Coherence \ac{PDF} $P(\Delta\vec{\ln D_\mathrm{eff}}, \vec{\Delta t}, \vec{\Delta \phi}\mid ...)$ for the LIGO Hanford (H1) and Livingston (L1) detector pair, shown in the $(\vec{\Delta t}, \vec{\Delta \phi})$ plane at a fixed value of $\Delta\vec{\ln \mathcal{D}}\approx 0.15$. Left: Wide single-pixel coherence \ac{PDF} for a targeted search using SNR $1.25, 1.75$ for LIGO Hanford and Livingston, computed for a specific sky location ($P(\hat{\Omega})=\delta(\hat{\Omega}-\hat{\Omega}^\prime)$, corresponding to one \texttt{healpix} pixel with nside$=8$). Right: Coherence PDF of the same sky location using SNR $5,7$ for LIGO Hanford and Livingston. The bright spots in the single-pixel \ac{PDF} correspond to the physically allowed combinations of time delay and phase difference for that particular sky location. The width of the bright region(probability peak) are controlled by the reference SNR used. Color intensity represents the normalized probability density.}
\end{figure}

We adjust the width of the coherence \ac{PDF} by choosing representative \ac{SNR} values, $\rho$, for hypothetical signals in each detector. The conventional all-sky configuration and the targeted configuration shown in \figref{fig:dtdphi_compare} adopt representative \ac{SNR} values of 5 and 7 for \ac{LIGO} Hanford and Livingston, respectively, and 4 for Virgo. To investigate the impact of broader priors, we also computed the coherence \ac{PDF} using lower representative \ac{SNR} values of $\rho = 1.25$, 1.75, and 1 for \ac{LIGO} Hanford, Livingston, and Virgo, respectively, and used these in the injection campaign described in \secref{sec:results_locinj_rec}.
\figref{fig:dtdphi_compare_width} provides a visual comparison between the coherence \ac{PDF}s constructed with higher and lower representative \ac{SNR} values. The bright peak in the \ac{PDF} corresponding to the higher representative \ac{SNR} values is substantially narrower than that obtained with the lower representative \ac{SNR} values.

In addition to the test described in \secref{sec:results_locinj_setup}, we performed another targeted search using the coherence PDF with narrower width (i.e., higher representative \ac{SNR}). Table~\ref{tab:recovery} summarizes the number of recovered injections out of 9457 injections in total for each configuration.
While the all-sky search recovered 818, using the broad \ac{PDF}, the pipeline detected 859 injections, which is 41 more than the all-sky search. This corresponds to an $\sim5\%$ increase in the number of injections recovered as discussed in \secref{sec:results_locinj_rec}. However, the run with narrower coherence PDF recovered $795$ injections, which is even $\sim3\%$ less than the allsky run. 
\begin{table}[h]
    \caption{Injection recovery counts for configurations where different coherence PDF used. There were 9457 simulated \ac{BNS} signals in the input data. The targeted search with a broad targeted PDF recovers the most injections, about 5\% more than the all-sky search. The narrow prior yields a recovery count slightly lower than the all-sky case, indicating that an overly tight prior can miss some true signals.}
    \label{tab:recovery}
    \centering
    \begin{tabular}{|l|c|}
        \hline
        \textbf{Search configuration} & \textbf{Recovered injections}\\
        \hline
        \hline
        All-sky            & 818 \\
        \hline
        Targeted (narrow PDF)           & 795 \\
        \hline
        Targeted (broad PDF)           & 859 \\
        \hline
    \end{tabular}
\end{table}
 
These results suggest that the coherence \ac{PDF} that is excessively narrow can degrade search sensitivity. Examining the missed injections in that case revealed that in several instances the real signal's parameters led to a slight mismatch in arrival time or phase, which falls in the low probability region off of the narrow peak. Thus those injections present were assigned a lower ranking statistic, while the broad coherence \ac{PDF} can encompass these small deviations more robustly. Importantly, the broad coherence \ac{PDF} still concentrates enough on the target location that it boosts the signals of interest relative to background. The fact that it recovered more than the allsky configuration demonstrates that even a moderately informative coherence PDF can improve sensitivity.

\section{Performance improvement in targeted searches for various sky locations}
\label{app:multi_locations}

To verify that the sensitivity improvements demonstrated in \secref{sec:results_locinj_rec} for the GW170817 sky location are representative of targeted searches more generally, we conducted additional localized injection campaigns at five other sky positions distributed across different regions of the sky. These positions were chosen to sample various declinations and antenna patterns of the detector network, ensuring that our results are not specific to a particular geometric configuration. Each campaign followed the same methodology as described in \secref{sec:results_locinj_rec}: 9457 \ac{BNS} injections uniformly distributed in redshift (up to $z=0.15$) were simulated at a fixed celestial coordinate, and the data were processed with both all-sky and targeted coherence \acp{PDF}.

\begin{figure*}[htbp]
    \centering
    \begin{minipage}[t]{0.48\textwidth}
        \centering
        \vspace{-3.5cm}
        \begin{tabular}{|l|c|c|c|}
            \hline
            \textbf{Sky Location} & \textbf{RA} & \textbf{Dec} & \textbf{VT Ratio}\\
            \hline
            \hline
            GW170817 (NGC 4993) & $13^{\rm h}09^{\rm m}$ & -23°22' & 1.08 \\
            \hline
            \texttt{HealPix} ID 6 & $07^{\rm h}30^{\rm m}$ & 78°17' & 1.12 \\
            \hline
            \texttt{HealPix} ID 33 & $14^{\rm h}15^{\rm m}$ & 66°26' & 1.09 \\
            \hline
            \texttt{HealPix} ID 100 & $14^{\rm h}08^{\rm m}$ & 48°08' & 1.11 \\
            \hline
            \texttt{HealPix} ID 121 & $07^{\rm h}07^{\rm m}$  & 41°48' & 1.08 \\
            \hline
            \texttt{HealPix} ID 377 & $07^{\rm h}07^{\rm m}$  & 00°00'& 1.05 \\
            \hline
        \end{tabular}
        \vspace{0.2cm}
        \captionof{table}{\ac{VT} improvements for targeted searches at six different sky locations. The \ac{VT} ratio represents the targeted search \ac{VT} divided by the all-sky search \ac{VT} at \ac{FAR} = 1/month. All positions show consistent improvements in the range of 5-12\%.}
        \label{tab:multi_location_vt}
    \end{minipage}
    \hfill
    \begin{minipage}[t]{0.48\textwidth}
        \centering
        \includegraphics[width=\linewidth]{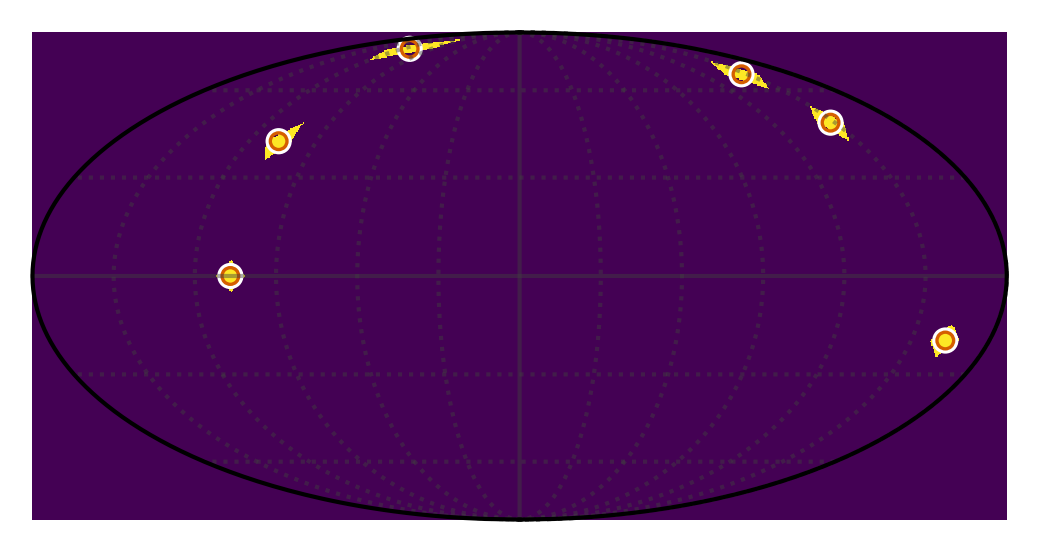}
        \caption{HEALPix skymap (nside=8) showing the sky locations of the six injection campaigns used to validate the targeted search methodology. Yellow pixels indicate the specific sky positions where injections were simulated.}
        \label{fig:healpix_skymap}
    \end{minipage}
\end{figure*}

\tabref{tab:multi_location_vt} summarizes the \ac{VT} improvements at \ac{FAR} = 1/month across all six sky locations tested, including GW170817's. The targeted search consistently outperforms the all-sky search across all positions, with \ac{VT} improvements ranging from approximately 5\% to 12\%. The variation in improvement magnitude can be attributed to differences in the network antenna pattern sensitivity and intrinsic uncertainties in the sensitivity measurement.
Nevertheless, all tested locations demonstrate meaningful sensitivity enhancements, confirming the robustness of the targeted search methodology.

It is worth noting that the optimal width of the coherence \ac{PDF} (controlled by the representative \ac{SNR} values, as discussed in \appref{app:pdf_width}) appears to be relatively insensitive to sky location. We used the same broad targeted \ac{PDF} configuration (with representative \acp{SNR} of 1.25, 1.75, and 1 for LIGO Hanford, Livingston, and Virgo, respectively) for all five positions, and all achieved comparable improvements over their respective all-sky baselines. This suggests that the targeted search framework is well-suited for operational deployment, where a single set of precomputed coherence \acp{PDF} can be applied to follow up \ac{EM} triggers from any direction on the sky without requiring location-specific tuning.
These multi-location studies provide confidence that the targeted search methodology will perform reliably for real multimessenger follow-up scenarios during future observing runs, regardless of where in the sky an \ac{EM} counterpart is detected.

\bibliography{references}{}

\bibliographystyle{aasjournal}
\end{document}